\documentclass[prd,aps,twocolumn,showpacs,floats,floatfix]{revtex4}
\usepackage[dvips]{graphicx}
 
\begin{document}
 
\title{Constraining crystalline color superconducting quark matter with
gravitational-wave data }

\author{Lap-Ming Lin}
\email{lmlin@phy.cuhk.edu.hk}
\affiliation{Department of Physics and Institute of Theoretical Physics,
The Chinese University of Hong Kong, Hong Kong, China}
 
\date{October 15, 2007}
 
\begin{abstract} 
We estimate the maximum equatorial ellipticity sustainable by  
compact stars composed of crystalline color-superconducting quark matter. 
For the theoretically allowed range of the gap parameter $\Delta$, 
the maximum ellipticity could be as large as $10^{-2}$, which is about 4 
orders of magnitude larger than the tightest upper limit obtained by the 
recent science runs of the LIGO and GEO600 gravitational wave detectors 
based on the data from 78 radio pulsars. 
We point out that the current gravitational-wave strain upper limit
already has some implications for the gap parameter. In particular, the upper
limit for the Crab pulsar implies that $\Delta$ is less than $O(20)$ MeV
for a range of quark chemical potential accessible in compact stars,
assuming that the pulsar has a mass $1.4 M_{\odot}$, radius $10$ km,
breaking strain $10^{-3}$, and that it has the maximum quadrupole
deformation it can sustain without fracturing.
\end{abstract}


\maketitle

\paragraph*{Introduction.}

When nuclear matter is squeezed to a sufficiently high density, there is 
a transition from nuclear matter to quark matter. Since the density required
for the transition to happen is believed to be not much higher than 
nuclear-matter density, the dense cores of compact stars are
the most likely places where quark matter may occur astrophysically. 
Except for hot newborn compact stars, it is now generally believed that the 
deconfined quark matter (if exists) in the interior of compact stars is in a 
color-superconducting phase \cite{alf98,rap98,alf99,alf03}, in which the 
quarks form Cooper pairs due to the BCS mechanism 
(see \cite{alf06} and references therein for reviews). 
At sufficiently high densities, the favored pairing phase is the 
color-flavor-locked (CFL) phase \cite{alf99}, in which pairing between 
quarks of different colors and flavors is allowed. 
In the intermediate density regime relevant to the cores of compact
stars, it is found that crystalline color-superconducting quark matter 
is a more favored phase \cite{alf01,alf06,cas05,man06,raj06,cas06}. 
However, it should be noted that so far the studies on the crystalline 
phase are based only on phenomenological models of QCD, mainly the 
Nambu-Jona-Lasinio model \cite{nam61}. 
The true ground state of quark matter in the cores of compacts stars
is still a matter of debate. 
Furthermore, there is as yet no study on the construction and stability
of a hydrostatic equilibrium stellar model composed of crystalline 
quark core in general relativity.

While it is still an open question whether quark stars exist in nature, it 
is interesting to ask how their existence would affect observable phenomena 
related to compact stars. Furthermore, could we make use of observation data 
to constrain various theoretical models? In this paper, we show that 
the observational upper limits on gravitational-wave emission from known 
pulsars can be used to constrain compact star models 
composed of crystalline color superconducting quark matter. The prospect of 
detecting the gravitational-wave signals emitted by quark stars has been 
considered before (e.g., \cite{che98,mar02,oec04,yas05,lin06}). 
However, our work represents the first attempt
to make use of real observational data to set constraint on theoretical
parameters of the quark matter.

Rotating compact stars are among the most promising  
gravitational-wave sources for Earth-based interferometric detectors such as 
LIGO, VIRGO, GEO600, and TAMA300. 
These objects can emit gravitational waves if they are asymmetric about the 
rotation axis. A number of mechanisms to give rise to the asymmetry have been 
proposed, including (1) nonaxisymmetric distortions of the solid crust or
core of the star \cite{bil98,ush00,owe05,has06}; (2) rotational induced 
instabilities such as the bar-mode and $r$-mode instabilities 
(see \cite{and01,and03} for reviews); 
(3) distortion produced by strong magnetic fields \cite{bon96,cut02a,mel05}; 
and (4) free precession of the star \cite{cut00,jon02,van05}. 
Here we study the nonaxisymmetric distortions of solid compact 
stars composed of crystalline color-superconducting quark matter 
and show that the recent observational results by LIGO and GEO600 
are already sensitive enough to put constraints on the gap parameter 
$\Delta$ associated with the color-superconducting phase. 
 
Over the past few years, the LIGO and GEO600 detectors have conducted 
four science runs starting in 2002. In particular, data from the second 
science run (denoted as S2) have been used to set direct upper limits on the
gravitational-wave amplitudes from 28 known pulsars \cite{abb05}. 
The lowest gravitational-wave strain upper limit in the S2 run is at the 
level $\sim 10^{-24}$ and the equatorial ellipticities are $\sim 10^{-5}$. 
Recently, the LIGO Science Collaboration has presented the latest results
for 78 pulsars based on data from the third and fourth science 
runs (denoted as S3 and S4) \cite{abb07}. The improved sensitivity
of the detectors in the S3/S4 runs leads to the tightest strain upper limit 
$\sim 10^{-25}$ and equatorial ellipticity $\sim 10^{-6}$. 

For conventional neutron star models, the maximum ellipticity 
that can be supported by the neutron star's solid crust is about 
$10^{-7}$ \cite{ush00,owe05}. The current detectors are still not sensitive 
enough to probe the upper range of ellipticity permitted by neutron stars. 
However, the current limits are well into the range permitted by some exotic
compact star models. 
Based on the hypothesis proposed by Xu \cite{xu03} that quarks might form 
clusters via a Van der Waals-type interaction, Owen \cite{owe05} 
found that solid compact stars composed of such quark-clusters 
might sustain ellipticity up to $\sim 10^{-4}$. It should be noted that 
the solid quark-star models considered in this paper, namely, the crystalline 
color-superconducting quark matter, is more robust than Xu's hypothesis 
from a theoretical point of view. 
We find that the maximum ellipticity sustainable by these models 
could be as large as $5\times 10^{-2}$. This makes the current strain upper 
limits obtained by the LIGO and GEO600 detectors become more astrophysically 
relevant and interesting.

\paragraph*{Elastic deformations and gravitational-wave emission.}

A triaxial pulsar, rotating about a principal axis, would radiate 
gravitational waves at twice the rotation frequency. In the quadrupole
approximation, the characteristic strain amplitude is \cite{jar98}
\begin{equation}
h_0 = {16\pi^2 G\over c^4} { I_{zz} f^2\over r} \epsilon , 
\label{eq:wave_h0}
\end{equation}
where $f$ is the pulsar's spin frequency, $r$ is the distance to the pulsar,
$\epsilon = (I_{xx}-I_{yy})/I_{zz}$ is the equatorial ellipticity, and 
$I_{ij}$ is the moment of inertia tensor. The $z$ axis is the rotation axis.
Using the definition of the mass multipole moment, 
$Q_{lm} = \int \rho r^l Y_{lm}^* d^3 x$, it can be shown that 
the ellipticity is related to the $m=2$ quadrupole moment $Q_{22}$ 
by 
\begin{equation}
\epsilon = \sqrt{32\pi\over 15} {Q_{22}\over I_{zz} } .
\label{eq:epsilon}
\end{equation}

In this paper, the pulsar's quadrupole moment is assumed to be due to its 
elastic deformation in the presence of a solid core. Assuming that the 
material is everywhere strained to the maximum, the 
maximum quadrupole moment the star can sustain is given by 
Ushomirsky {\it et al.} \cite{ush00}:  
\begin{equation}
Q_{\rm max}  = \sqrt{32\pi\over 15} \sigma_{\rm max} 
\int {\nu r^3\over g} \left( 48 - 14 \tilde{U} + \tilde{U}^2 
- r {  d\tilde{U}\over dr } \right) dr , 
\label{eq:Q22}
\end{equation}
where $\sigma_{\rm max}$ is the breaking strain, $\nu$ is the shear modulus,
$g=G M(r)/r^2$ is the local gravitational acceleration, and 
$\tilde{U} = 2 + d \ln g / d \ln r$. 
For a uniform-density incompressible model, which is a good 
approximation to quark stars considered in this paper,  
$Q_{\rm max}$ is simplified to  
\begin{equation}
Q_{\rm max} \approx { 13 \nu \sigma_{\rm max} R^6\over G M } . 
\label{eq:Qmax}
\end{equation}
It should be noted that in deriving Eq.~(\ref{eq:Q22}), the self-gravity of 
the deformation is neglected. 
However, dropping that approximation would only change $Q_{\rm max}$ by 
a factor of 2 \cite{has06}. This is not large enough to change the 
implications we shall draw from the estimate of the ellipticity described 
below. 

It is seen from Eq.~(\ref{eq:Q22}) that more rigid material (i.e., material 
with a larger shear modulus $\nu$) can lead to a larger quadrupole 
deformation, which in turn implies a larger ellipticity 
[see Eq.~(\ref{eq:epsilon})]. 
This is the main reason why solid quark stars can sustain much larger 
ellipticity than conventional neutron stars since the shear modulus of 
solid quark matter can be a few orders of magnitude larger than that of the 
neutron star's crust \cite{owe05} (also see below). 
In Eq.~(\ref{eq:Q22}) we define the shear modulus to be half the ratio of 
the stress to the strain, which has a factor of 2 different from that 
defined in the original paper by Ushomirsky {\it et al.} \cite{ush00}. 
This definition is consistent with that used by Mannarelli {\it et al.} 
\cite{man07} in calculating the shear modulus of crystalline 
color-superconducting quark matter, which is given by 
\begin{equation}
\nu = 2.47\ {\rm MeV/fm^{3} } \left( {\Delta\over 10\ {\rm MeV} }\right)^2
\left( {\mu\over 400\ {\rm MeV} }\right)^2 ,
\label{eq:nu_CQM}
\end{equation}
where $\Delta$ is the gap parameter and $\mu$ is the quark chemical 
potential. 
We remark that this result is obtained by performing a Ginzburg-Landau
expansion to order $\Delta^2$. Since the control parameter for the expansion
is $[ \Delta/(m_s^2/8\mu) ] \sim 1/2$ (with $m_s$ being the strange quark 
mass) for the favored crystalline phase \cite{man07}, 
Eq.~(\ref{eq:nu_CQM}) can only be used to fix the order of magnitude of $\nu$. 
Higher-order correction terms are suppressed only by about $1/4$.

To describe quark matter within compact stars, 
the relevant range of $\mu$ is 
$350\ {\rm MeV} < \mu < 500\ {\rm MeV}$ \cite{ipp07,man07}.  
In the lower region of this window, Ippolito {\it et al.} 
\cite{ipp07} notice that the crystalline phase is more likely to be a 
two-flavor ($\langle ud \rangle$), rather than the three-flavor 
($\langle ud \rangle$ and $\langle us \rangle$) 
condensate as considered in \cite{man07}.
In the CFL phase, where the density is high enough so that the effects of 
the strange quark mass can be neglected, the gap parameter $\Delta$ is 
expected to lie between 10 and 100 MeV \cite{alf06,man07}. 
However, for the quark matter to be in the crystalline phase 
rather than the CFL phase, Mannarelli {\it et al.} \cite{man07}
estimate that the reasonable range for $\Delta$ is
$5\ {\rm MeV} \lesssim \Delta \lesssim 25\ {\rm MeV}$.
These windows for $\mu$ and $\Delta$ imply that the 
shear modulus of crystalline color-superconducting quark matter within 
compact stars is in the range
$0.47\ {\rm MeV/fm^3} < \nu < 24\ {\rm MeV/fm^3 }$.
For comparison, the shear modulus of the neutron star's crust is of the order
$1\ {\rm keV/fm^3}$ \cite{ush00,man07}. 

The breaking strain $\sigma_{\rm max}$ is highly uncertain. This is the value 
beyond which solid materials no longer behave elastically. Instead, they would 
either crack or undergo plastic deformation. For perfect crystals without 
defects, $\sigma_{\rm max}$ could be as high as $10^{-1}$ \cite{kit95}. 
However, it is also known that $\sigma_{\rm max}$ of real crystals can be 
several orders of magnitude lower. The discrepancy is explained by the 
presence of defects (e.g., dislocations) in real crystals. 
Based on the extrapolation of terrestrial materials, values as high as 
$10^{-2}$ has been suggested for neutron star crusts \cite{rud91,rud92}, 
which may also be favored by the enormous energy ($\sim 10^{46}\ {\rm erg}$) 
liberated in the December 27, 2004 giant flare of SGR 1806-20 
according to the magnetar model \cite{tho95,hur05}. 
In this work, we consider the 
effects of $\sigma_{\rm max}$ in the range $10^{-3}$ to $10^{-2}$, values 
that have been used in the study of compact stars by other authors 
(e.g., \cite{tho95,ush00,jon02,owe05}).

Using the shear modulus given in Eq.~(\ref{eq:nu_CQM}), the maximum 
quadrupole moment for a solid quark star in the crystalline 
color-superconducting phase can be estimated by Eq.~(\ref{eq:Qmax}). However, 
the moment of inertia $I_{zz}$ is needed to calculate the corresponding 
maximum equatorial ellipticity. Using the empirical formula for strange 
stars given by Bejger and Haensel (see Eq.~(10) of \cite{bej02}), 
Eq.~(\ref{eq:epsilon}) gives the maximum equatorial ellipticity as 
\begin{eqnarray}
&&\epsilon_{\rm max} = 2.6\times 10^{-4}
\left( {\nu\over {\rm MeV/fm^3} } \right)
\left( {\sigma_{\rm max}\over 10^{-3}} \right)
\left( {1.4M_\odot\over M} \right)^2  \cr
&& \cr
&& \times \left( {R\over 10\ {\rm km}} \right)^4 
\left[ 1 + 0.14\left( {M\over 1.4M_\odot} \right)
\left( {10\ {\rm km}\over R} \right) \right]^{-1} .
\end{eqnarray}
We use the values $M=1.4 M_\odot$, $R=10$ km, and 
$\sigma_{\rm max}=10^{-2}$ in order to compare with \cite{owe05}, in which 
Owen obtained $\epsilon_{\rm max}\sim 2\times 10^{-4}$ for solid strange stars 
(with the quarks clustered in groups of about 18). For the estimated range of 
the shear modulus given above, we find that 
$\epsilon_{\rm max}$ could be as large as $\sim 5\times 10^{-2}$ for 
solid quark 
stars in a crystalline color-superconducting phase. This relatively large 
value of $\epsilon_{\rm max}$ is about 4 orders of magnitude larger than 
the tightest upper limit obtained by the combined S3/S4 result for the 
pulsar PSR J2124-3358 \cite{abb07}.

\begin{figure}
\centering
\includegraphics*[width=7cm]{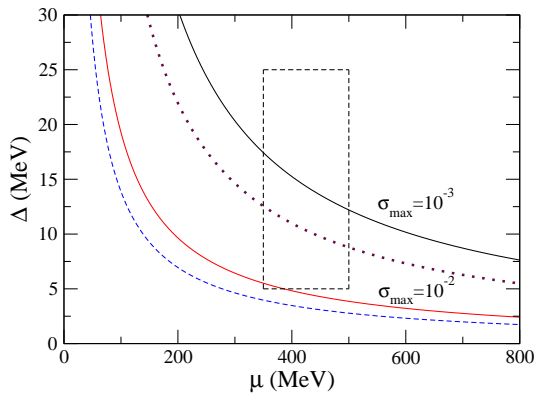}
\caption{(color online). 
The areas above the solid lines are excluded by the direct  
upper limit for the Crab pulsar obtained from the S3/S4 runs for the 
cases $\sigma_{\rm max}=10^{-3}$ and $10^{-2}$. The dotted 
(dashed) line is the constraint set by the spin-down limit for 
$\sigma_{\rm max}=10^{-3}$ ($10^{-2}$). The rectangular box is the 
theoretically allowed region of $\mu$ and $\Delta$. }
\label{fig:fixMR}
\end{figure}

\paragraph*{Constraints set by the Crab pulsar.}

Eq.~(\ref{eq:wave_h0}) suggests that the observational upper limits 
on $h_0$ obtained 
from known isolated pulsars can be used to set a limit on $\epsilon$ 
assuming a value of $I_{zz}$. However, the moment of inertia is very 
sensitive to the poorly known dense matter equation of state (EOS). It can 
change by a factor of 7 depending on the stiffness of the EOS 
\cite{bej02}. Alternatively, with Eq.~(\ref{eq:epsilon}), one can use 
Eq.~(\ref{eq:wave_h0}) to set a limit on the pulsar's quadrupole moment 
without assuming a value of $I_{zz}$ \cite{pit05}. 
The limit can in turn set a constraint on the shear modulus of crystalline
color-superconducting quark matter by Eq.~(\ref{eq:Qmax}). In particular, 
with the expression~(\ref{eq:nu_CQM}) for $\nu$, we can 
define an exclusion region in the $\Delta-\mu$ plane by the the following 
constraint:
\begin{eqnarray}
&& \Delta \mu \lesssim 7.3\times 10^4\ {\rm MeV^2}
\left( {10\ {\rm km}\over R} \right)^3
\left( { 1\ {\rm Hz}\over f} \right)  \cr
&& \cr
&& \times 
\left[ 
\left({\tilde{h}_0\over 10^{-24} }\right) 
\left( {M\over 1.4M_{\odot}}\right)
\left( {10^{-3}\over \sigma_{\rm max} } \right)
\left( {r\over 1\ {\rm kpc}} \right)
\right]^{1/2}  , 
\label{eq:Delta_mu} 
\end{eqnarray}
where $\tilde{h}_0$ is the observational upper limit on $h_0$ for a given 
pulsar. 

Under the assumption that the pulsar is an isolated rigid body and that the
observed spin-down of the pulsar is due to the loss of rotational kinetic 
energy as gravitational radiation, one can also obtain the so-called spin-down 
limit on the gravitational-wave amplitude 
$h_{\rm sd} = ( {5 G I_{zz} |\dot{f}| / 2 c^3 r^2 f } )^{1/2}$,
where $\dot{f}$ is the time derivative of the pulsar's spin frequency 
\cite{abb07}. As it is expected that the strain amplitude satisfies 
$h_0 \lesssim h_{\rm sd}$ in general, we can 
derive a constraint on the product $\Delta \mu$ based on the spin-down limit: 
\begin{eqnarray}
\left( \Delta\mu \right)_{\rm sd} &\lesssim& 
2.1\times 10^5\ {\rm MeV^2}
\left( {10^{-3}\over \sigma_{\rm max}} \right)^{1/2} 
\left( {M\over 1.4M_\odot} \right)^{3/4} \cr 
&& \cr
&& 
\times \left( {10\ {\rm km}\over R} \right)^{5/2} 
\left( { |\dot{f}|\over 10^{-10}\ {\rm Hz\ s^{-1}} } \right)^{1/4}
\left( {f\over {\rm 1\ Hz}} \right)^{-5/4} \cr
&& \cr
&& \times \left[ 1 + 0.14\left({M\over 1.4M_\odot} \right)
\left( {10\ {\rm km}\over R} \right) \right]^{1/4} , 
\label{eq:Delta_mu_sd}
\end{eqnarray} 
where we have used Eq.~(10) of \cite{bej02} for the moment of inertia 
for strange stars.

In \cite{abb07} it is reported that the gravitational-wave strain upper limit 
for the Crab pulsar ($\tilde{h}_0=3.1\times 10^{-24}$) is the closest to the 
spin-down limit (at a ratio of 2.2). 
For the other pulsars, the direct observational upper 
limits are typically at least 100 times larger than the spin-down 
limits. 
For pulsars in globular clusters, in which cases the spin-down 
measurement is obscured by the cluster's dynamics, the gravitational-wave 
observations provide the only direct upper limits. 
In the following, we shall focus on the constraints set by the observational 
data for the Crab pulsar. 

\begin{figure}
\centering
\includegraphics*[width=7cm]{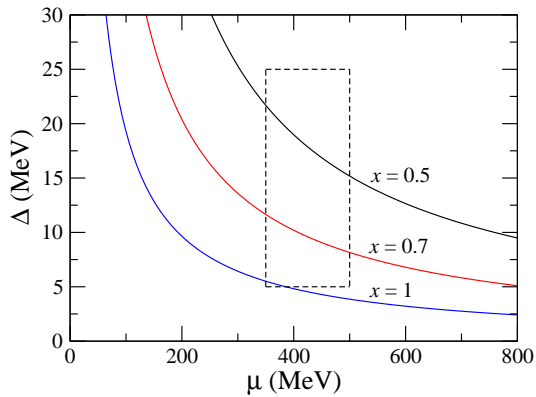}
\caption{(color online). 
The dependence of the exclusion region on the size of the 
crystalline core. The mass, radius, and breaking strain are fixed at 
$M=1.4M_\odot$, $R=10$ km, and $\sigma_{\rm max}=10^{-2}$. The parameter
$x=R_{\rm c}/R$, where $R_{\rm c}$ is the radius of the core.} 
\label{fig:vary_x}
\end{figure}

The S3/S4 results for the Crab pulsar are plotted in Fig.~\ref{fig:fixMR}.
In the figure, the solid lines represent Eq.~(\ref{eq:Delta_mu}) when the 
equality holds for the breaking strain $\sigma_{\rm max}=10^{-3}$ and 
$10^{-2}$, assuming that $M=1.4M_\odot$ and $R=10\ {\rm km}$. 
The pulsar's spin frequency is $f=29.8\ {\rm Hz}$ and the distance 
is $r=2\ {\rm kpc}$. 
For comparison, the dotted and dashed lines are the constraints 
obtained by the spin-down limit [Eq.~(\ref{eq:Delta_mu_sd})], respectively, 
for $\sigma_{\rm max}=10^{-3}$ and $10^{-2}$, 
with the pulsar's spin-down rate 
$|\dot{f}| = 3.73\times 10^{-10}\ {\rm Hz\ s^{-1}}$. In the figure, the region
to the right of each curve is excluded by the corresponding constraint. 
The rectangular box (in dashed line) in Fig.~\ref{fig:fixMR} encloses the 
theoretically allowed ranges of $\mu$ and $\Delta$ in compact stars as
discussed above. 
It is seen that the direct observational upper limits (solid lines) can 
already place a strong constraint on the possibility that the Crab pulsar 
might be a solid star composed of crystalline color superconducting quark 
matter. For the fiducial values $M=1.4M_\odot$, $R=10$ km, and 
$\sigma_{\rm max}=10^{-3}$, the gap parameter $\Delta$ is restricted to be 
less than $\sim O(20)$ MeV for the range of quark chemical potential relevant 
to compact stars. Furthermore, the more extreme estimate
$\sigma_{\rm max}=10^{-2}$ excludes essentially the entire theoretically 
allowed range of $\Delta$.

So far we have assumed that the whole star is 
composed of crystalline quark matter. However, a more realistic compact star
model would be a crystalline quark-matter core surrounded by a nuclear-matter
fluid layer, which is then followed a thin crust. As a first step to 
investigate the effects of a smaller crystalline core on the constraint
(\ref{eq:Delta_mu}), we assume that the quadrupole deformation 
(and hence the gravitational wave emission) is due 
only to the crystalline core, without considering the respond of the outer 
fluid layer and thin crust. The maximum quadrupole moment (\ref{eq:Qmax})
is replaced by 
$Q_{\rm max}\approx 13 \nu \sigma_{\rm max} R_{\rm c}^6/GM_{\rm c}$, 
where $M_{\rm c}$ and $R_{\rm c}$ are, respectively, the mass and radius of 
the crystalline core. To relate $Q_{\rm max}$ to the global parameters 
$M$ and $R$ of the star, it is necessary to know the mass distribution 
of the outer fluid layer, which depends on the nuclear-matter EOS.
For simplicity, we assume that the density falls off as 
$\rho(r) = \rho_{\rm c} [1-(r/R)^2]/[1-(R_{\rm c}/R)^2]$ in the outer 
fluid layer (i.e., for $R_{\rm c} \le r \le R$), where $\rho_{\rm c}$ is the 
density of the core \cite{footnote2}. 
The quadrupole moment can then be expressed as  
$Q_{\rm max} \approx  { 13 \nu \sigma_{\rm max} R^6 \alpha(x)/ GM }$,
where $x=R_{\rm c}/R$, 
$\alpha(x) = x^6 [ 1 - (1-x^2)^{-1}(1- 3x^2/5 - 2x^{-3}/5 ) ]$, and 
$\alpha(1) \equiv 1$. Accordingly, the constraint (\ref{eq:Delta_mu}) is 
modified by multiplying a factor $\alpha(x)^{-1/2}$ on the right hand side
of the equation. 
The results for three different values of $x$ are plotted 
in Fig.~\ref{fig:vary_x}, with $M=1.4M_\odot$, $R=10$ km, and 
$\sigma_{\rm max}=10^{-2}$ being fixed. 
The figure shows that a smaller crystalline core would make the extreme 
estimate $\sigma_{\rm max}=10^{-2}$ more compatible with the current upper 
limit for the Crab pulsar. 
However, the effect due to the respond 
of the outer fluid layer to the internal deformation deserves further 
investigation.

\paragraph*{Summary and discussion.}

In this paper, we point out that the S3/S4 runs of the LIGO and GEO 600 
gravitational-wave detectors are already sensitive enough to put an 
interesting constraint on solid quark-star models composed of crystalline 
color superconducting quark matter.  
We have estimated that the maximum equatorial ellipticity 
sustainable by these stellar models could be as large as $\sim 10^{-2}$.  
This is about 2 orders of magnitude larger than that obtained by Owen 
\cite{owe05} for solid quark star models composed of quark-clusters, without
taking into account the effects of color superconductivity. 
We have used the direct observational gravitational-wave strain upper 
limit for the Crab pulsar in the S3/S4 runs to constrain the poorly known 
gap parameter $\Delta$ in the crystalline color-superconducting phase. 
For $\sigma_{\rm max}=10^{-3}$, we conclude that the Crab pulsar could be 
made entirely of crystalline quark matter if $\Delta$ is less than 
$O(20)$ MeV. For the extreme estimate $\sigma_{\rm max}=10^{-2}$, the Crab 
pulsar is unlikely to be a complete solid quark star. However, it could 
still contain a smaller crystalline quark core in this case. 
The direct observational upper limit for the Crab pulsar should beat the 
spin-down limit in the fifth science run of the LIGO detectors \cite{abb07}. 
This promises to put a stronger constraint on the theoretical models
considered in this paper.

Finally, we conclude with a few remarks. 
(1) It should be noted that the existence of crystalline quark matter 
inside compact stars is still a matter of debate. As mentioned before, 
this phase of quark matter has only been studied in phenomenological models 
of QCD and no analysis on the gravitational stability of such stellar 
configuration has been carried out. 
Our work suggests a way to put constraint on such exotic compact
star models based on gravitational-wave observation. We also note that 
Anglani {\it et al.} \cite{ang06} have recently studied the cooling rate of 
these compact stars. This can provide another observational constraint.  
(2) A general argument against 
the existence of quark stars is that (fluid) quark-star models are inconsistent
to the behavior of pulsar glitches \cite{alp87}. The standard explanation for 
glitches in the conventional neutron star model involves the pinning and 
unpinning of large numbers of superfluid vortices to the solid crust 
\cite{and75,alp84,alp84a}. 
But the significant development of the effective theory of QCD 
over the past few years suggests that the quark matter inside compact 
stars is likely to be simultaneously superfluids and rigid solids, which are
the two crucial conditions for the standard model of glitches 
(see \cite{alf01,man07} for discussion and \cite{ang07} for the
study of the superfluid mode in the crystalline quark-matter phase). 
(3) One uncertainty in our estimate is the 
value of the breaking strain $\sigma_{\rm max}$. One should keep in mind that 
the values $\sigma_{\rm max}=10^{-3}-10^{-2}$ used in this paper probably 
lie in the upper end of the theoretical range \cite{rud91,rud92}. 
We also assume that the stellar material is everywhere strained to the 
maximum in the analysis. As pointed out by Owen \cite{owe05}, since a given 
pulsar may not be strained to the maximum, thus no upper limit 
on $\epsilon_{\rm max}$ can ever rule out any theoretical model. But the 
credibility of the model will face serious challenge as the number of 
observed pulsars increases and tighter limits on $\epsilon_{\rm max}$ are 
placed. 

The author thanks Pui Tang Leung for careful reading of the manuscript. 
This work is supported in part by the Hong Kong Research
Grants Council (grant no: 401905 and 401807) and the Chinese University of 
Hong Kong. 

\bibliographystyle{prsty}

\begin{thebibliography}{1}
\bibitem{footnote2}This is essentially the so-called 
Tolman VII model (apart from the normalization constant to make the 
density continuous across the interface), which can describe the density 
profile of a normal neutron star very well \cite{lat05}.
\end{thebibliography}

\end{document}